\documentclass[aps,showpacs,preprintnumbers,amsmath, amssymb]{revtex4}

\usepackage{float}
\usepackage{graphics,epsfig}
\usepackage{graphicx}
\usepackage{dcolumn}
\usepackage{bm}

\begin{document}
\baselineskip=0.8 cm

\title{{\bf Static scalar field condensation in regular asymptotically AdS reflecting star backgrounds}}
\author{Yan Peng$^{1}$\footnote{yanpengphy@163.com}}
\affiliation{\\$^{1}$ School of Mathematical Sciences, Qufu Normal University, Qufu, Shandong 273165, China}

\vspace*{0.2cm}
\begin{abstract}
\baselineskip=0.6 cm
\begin{center}
{\bf Abstract}
\end{center}

We study condensation behaviors of static scalar fields in the
regular asymptotically AdS reflecting star spacetime.
With analytical methods, we provide upper bounds for the radii of the scalar hairy reflecting stars.
Above the bound, there is no scalar hair theorem for the star.
Below the bound, we numerically obtain charged scalar hairy reflecting star solutions
and in particular, the radii of the hairy stars are discrete,
which is similar to known results in other reflecting object backgrounds.
For every set of parameters, we search for the largest AdS hairy star radius,
study effects of parameters on the largest hairy star radius
and also find difference between properties in this AdS reflecting star background
and those in the flat reflecting star spacetime.
Moreover, we show that scalar fields cannot condense around regular AdS
reflecting stars when the star charge is small or the cosmological constant is negative enough.

\end{abstract}

\maketitle
\newpage
\vspace*{0.2cm}

\section{Introduction}

The famous no-scalar-hair theorem announces that the static massive
scalar fields cannot exist in asymptotically flat black holes and
it was believed that this property is mainly due to the fact that a classical black hole horizon
could irreversibly absorb matter and radiation fields,
for references see\cite{Bekenstein}-\cite{Brihaye}
and a review see \cite{Bekenstein-1}.
In fact, this no scalar hair behavior is not restricted to
spacetime with a horizon.
Hod firstly found that regular asymptotically flat neutral compact reflecting
stars cannot support the static massive scalar fields \cite{Hod-6}.
This no hair theorem for regular
reflecting stars leads to works searching for no scalar hair
behavior in the horizonless spacetime.

As a further step, it was found that massless scalar fields nonminimally coupled to the gravity
cannot exist around the asymptotically flat neutral compact reflecting stars \cite{Hod-7}.
In addition, Bhattacharjee and Sudipta extended the discussion of no hair behaviors to
reflecting stars in the asymptotically dS gravity background \cite{Bhattacharjee}.
It seems that no hair theorem may widely hold in regular reflecting objects gravities.
It is known that there is no hair theorem for asymptotically flat charged black holes \cite{Hod-8,Hod-9}.
And it should be emphasized the fact that in Refs. \cite{Hod-8,Hod-9} the scalar field is
not necessarily static as usually assumed in various no-hair theorems.
Along this line, it is interesting to further examine whether scalar fields can exist
in the asymptotically flat charged reflecting objects background.
Very different from cases in black holes, it was lately found that
charged scalar fields can condense around a charged reflecting shell
and it was found that the radius of the hairy shell is discrete in a range,
where the spacetime outside the shell is supposed to be absolutely flat \cite{Hod-10,Hod-11}.
In fact, the existence of composed scalar fields and reflecting objects configurations
doesn't depend on the choice of flat spacetime limit.
In curved backgrounds of asymptotically flat reflecting stars, it was found that
the reflecting star can support massive scalar fields
and the scalar hairy reflecting star radius is also discrete \cite{Hod-12,YP2018}.

A simple way to invade the no hair theorem of static scalar fields is
enclosing the gravity system in a confined box \cite{SRD,Pallab Basu,YanP,YanP-1,Lu}.
And it is well known that the AdS boundary
could also serve as an infinity potential
to confine the scalar field and dynamical formations of scalar hair
due to the confinement was studied in \cite{PB}.
At present, AdS scalar hairy black holes have been widely studied with the
interest of the application of holographic theories \cite{Maldacena}-\cite{Yi Ling}.
On the other side, the scalar field configurations were constructed
in the regular AdS reflecting shell background with analytical methods \cite{YPS}.
For a AdS shell, we mean that the shell charge and mass is very small compared to the shell radius
or the spacetime outside the shell is pure AdS.
Along this line, it is interesting to extend the discussion to the more general
AdS reflecting star background by including
the nonzero star charge and mass to examine whether there are still no scalar hair behaviors
and also try to construct scalar field configurations
supported by AdS charged reflecting stars.

This paper is organized as follows. In section II,
we introduce the gravity model constructed with a
scalar field coupled to asymptotically AdS charged reflecting stars.
In part A of section III, we obtain an upper bound for the scalar hairy reflecting star radius.
And in part B of section III, we show that the scalar field could condense
around a AdS reflecting star and study properties of the largest hairy star radius.
The last section is a summary of our main results.

\section{Equations of motion and boundary conditions}

In this paper, we study the system of a scalar field coupled to a charged reflecting star
in the four dimensional asymptotically AdS spacetime.
And the corresponding Lagrange density reads
\begin{eqnarray}\label{lagrange-1}
\mathcal{L}=-\frac{1}{4}F^{MN}F_{MN}-|\nabla_{\mu} \psi-q A_{\mu}\psi|^{2}-m^{2}\psi^{2},
\end{eqnarray}
As usual we define $A_{\mu}$ as the ordinary Maxwell field with only nonzero tt component as $A_{\mu}=-\frac{Q}{r}dt$ and $\psi=\psi(r)$
is the scalar field with only radial dependence. Here, m is
the scalar field mass and q corresponds to the scalar field charge.

We set the ansatz of the asymptotically AdS star with planar symmetry as \cite{Hartnoll,YQ}
\begin{eqnarray}\label{AdSBH}
ds^{2}&=&-f(r)dt^{2}+\frac{dr^{2}}{f(r)}+r^{2}(dx^2+dy^{2}),
\end{eqnarray}
where $f(r)=\frac{r^2}{L^2}-\frac{2M}{r}+\frac{Q^2}{r^2}$ with $M$ as the star mass, $Q$ as the star charge
and L corresponding to the AdS radius.
The negative cosmological constant can be expressed with the AdS radius as $\Lambda=-\frac{3}{L^2}$ \cite{QBW}.
We also define the radial coordinate $r=r_{s}$ as the radius of the reflecting star.
In this work, we study the regular reflecting star without a horizon or there is
$g(r)>0$ for all $r\geqslant r_{s}$.

The scalar field equation can be obtained as
\begin{eqnarray}\label{BHg}
\psi''+(\frac{2}{r}+\frac{g'}{g})\psi'+(\frac{q^2Q^2}{r^2g^2}-\frac{m^2}{g})\psi=0,
\end{eqnarray}
where $g=\frac{r^2}{L^2}-\frac{2M}{r}+\frac{Q^2}{r^2}$.

We can set $L=1$ in the calculation with the symmetry
\begin{equation}
L\rightarrow \alpha L,~~r\rightarrow \alpha r,~~q\rightarrow q/\alpha,~~t\rightarrow \alpha t,~~M\rightarrow \alpha M,~~Q\rightarrow \alpha Q,~~m\rightarrow m/\alpha.
\end{equation}

Near the AdS boundary, asymptotic behaviors of the scalar fields are
$\psi=\frac{\psi_{-}}{r^{\lambda_{-}}}+\frac{\psi_{+}}{r^{\lambda_{+}}}+\cdots$
with $\lambda_{\pm}=(3\pm \sqrt{9+4m^2L^2})/2$ \cite{Horowitz,HCG}.
For $m^2L^2>0$, only the second scalar operator $\psi_{+}$ is normalizable.
So we fix $\psi_{-}=0$ and the scalar fields around the infinity behave as
\begin{equation}
\psi=\frac{\psi_{+}}{r^{\lambda_{+}}}+\cdots.
\end{equation}

We also impose reflecting boundary conditions
at the surface of the star as
\begin{equation}
\psi(r_{s})=0.
\end{equation}

\section{Scalar field condensation behaviors around AdS charged reflecting stars}

\subsection{Upper bounds for the radius of the scalar hairy AdS reflecting star}

With a new radial function $\tilde{\psi}=\sqrt{r}\psi$,
the equation of the scalar field can be expressed as
\begin{eqnarray}\label{BHg}
r^2\tilde{\psi}''+(r+\frac{r^2g'}{g})\tilde{\psi}'+(-\frac{1}{4}-\frac{rg'}{2g}+\frac{q^2Q^2}{g^2}-\frac{m^2r^2}{g})\tilde{\psi}=0,
\end{eqnarray}
with $g=\frac{r^2}{L^2}-\frac{2M}{r}+\frac{Q^2}{r^2}$.

From the boundary conditions (5) and (6), we have
\begin{eqnarray}\label{InfBH}
&&\tilde{\psi}(r_{s})=0,~~~~~~~~~\tilde{\psi}(\infty)=0.
\end{eqnarray}

The function $\tilde{\psi}$ must have (at least) one extremum point $r=r_{peak}$
between the surface $r_{s}$ of the reflecting star and the AdS boundary $r_{b}=\infty$.
At this extremum point, the scalar field is characterized by
\begin{eqnarray}\label{InfBH}
\{ \tilde{\psi}'=0~~~~and~~~~\tilde{\psi} \tilde{\psi}''\leqslant0\}~~~~for~~~~r=r_{peak}.
\end{eqnarray}

According to the relations (7) and (9), we arrive at the inequality
\begin{eqnarray}\label{BHg}
-\frac{1}{4}-\frac{rg'}{2g}+\frac{q^2Q^2}{g^2}-\frac{m^2r^2}{g}\geqslant0~~~for~~~r=r_{peak}.
\end{eqnarray}

Then we have
\begin{eqnarray}\label{BHg}
m^2r^2g\leqslant q^2Q^2-\frac{rgg'}{2}-\frac{1}{4}g^2~~~for~~~r=r_{peak}.
\end{eqnarray}

We divide radii of hairy reflecting stars into two types: $r_{s}< max\{\sqrt[3]{4ML^2},\sqrt{QL}\}$
and $r_{s}\geqslant max\{\sqrt[3]{4ML^2},\sqrt{QL}\}$.
For radii satisfying $r_{s} \leqslant  max \{\sqrt{QL},\sqrt[3]{4ML^2}\}$, we
naturally have an upper bound $m r_{s}\leqslant max \{m\sqrt{QL},m\sqrt[3]{4ML^2}\}$.
For other radii satisfying $r_{s}\geqslant max\{\sqrt[3]{4ML^2},\sqrt{QL}\}$,
we obtain an upper bound in the following.

For radii satisfying $r_{s}\geqslant max\{\sqrt[3]{4ML^2},\sqrt{QL}\}$, there is
\begin{eqnarray}\label{BHg}
g=\frac{r^2}{L^2}-\frac{2M}{r}+\frac{Q^2}{r^2}=\frac{r^2}{2L^2}+\frac{r^2}{2L^2}-\frac{2M}{r}+\frac{Q^2}{r^2}
=\frac{r^2}{2L^2}+\frac{1}{2rL^2}(r^3-4ML^2)+\frac{Q^2}{r^2}\geqslant\frac{r^2}{2L^2}
\end{eqnarray}
and
\begin{eqnarray}\label{BHg}
g'=(\frac{r^2}{L^2}-\frac{2M}{r}+\frac{Q^2}{r^2})'=\frac{2r}{L^2}+\frac{2M}{r^2}-\frac{2Q^2}{r^3}=\frac{2}{r^3L^2}(r^4-Q^2L^2)+\frac{2M}{r^2}\geqslant 0.
\end{eqnarray}

From the regular condition $g(r)>0$ for all $r>r_{s}$ and relations (11)-(13), we arrive at
\begin{eqnarray}\label{BHg}
m^2r_{s}^2\frac{r_{s}^2}{L^2}\leqslant m^2r^2\frac{r^2}{L^2}\leqslant m^2r^2\cdot 2g(r) \leqslant 2(q^2Q^2-\frac{rgg'}{2}-\frac{1}{4}g^2)\leqslant 2q^2Q^2~~~for~~~r=r_{peak}.
\end{eqnarray}

According to (14), there is
\begin{eqnarray}\label{BHg}
m^2r_{s}^2\frac{r_{s}^2}{L^2}\leqslant  2q^2Q^2.
\end{eqnarray}

It also can be expressed with dimensionless quantities with the symmetry (4) as
\begin{eqnarray}\label{BHg}
m r_{s}\leqslant  \sqrt[4]{2}\sqrt{qQmL}.
\end{eqnarray}

Then we have two types of hairy star radii

~~~~case 1:~~~$r_{s} \leqslant  max \{\sqrt{QL},\sqrt[3]{4ML^2}\}$~or~$m r_{s}\leqslant max \{m\sqrt{QL},m\sqrt[3]{4ML^2}\}$;

~~~~case 2:~~~$r_{s}\geqslant  max \{\sqrt{QL},\sqrt[3]{4ML^2}\}$~with~$m r_{s}\leqslant  \sqrt[4]{2}\sqrt{qQmL}$.

So we obtain an upper bound for the regular AdS hairy reflecting star radii as
\begin{eqnarray}\label{BHg}
m r_{s}\leqslant  max \{m\sqrt{QL},m\sqrt[3]{4ML^2},\sqrt[4]{2}\sqrt{qQmL}\}.
\end{eqnarray}

\subsection{Scalar field configurations supported by a regular AdS charged reflecting star}

In this part, we construct the regular AdS scalar hairy charged
reflecting star solutions and also study properties of the hairy reflecting star radii.
Since we impose reflecting boundary conditions for the scalar field
at the star surface, the scalar field can be putted in the form
$\psi=\psi_{0}(r-r_{s})+\cdots$ around the star radius.
As the scalar field equation is linear with respect to $\psi$,
we can fix $\psi_{0}=1$ in the calculation.
We will numerically integrate the equation from various reflecting boundary
to the infinity to search for hairy star radii $r_{s}$
satisfying the boundary conditions (5).

We show the case of $qL=2$, $m^2L^2=\frac{1}{2}$, $\frac{Q}{L}=4$ and $\frac{M}{L}=7.5$ in Fig. 1.
In the left panel, when we impose a reflecting boundary at $\frac{r}{L}=1.900$, the scalar field is
zero around $\frac{r}{L}\thickapprox3.061$ and it decreases to be more negative far away.
And with a little larger reflecting boundary at $\frac{r}{L}=1.909$
in the middle panel, the scalar field decreases to be zero at a larger coordinate $\frac{r}{L}\thickapprox5.297$
and becomes more negative far away from the reflecting boundary.
For a reflecting boundary at $\frac{r}{L}=1.918$ in the right panel,
there is no additional zero point of the scalar field
and the scalar field increases to be more positive far from the star.
In front cases, it is clearly that the reflecting boundaries cannot be fixed as the star radius
since the scalar field should asymptotically approach zero far from the star surface
according to (5).
However, there may exist a critical reflecting boundary between $\frac{r}{L}=1.909$ and $\frac{r}{L}=1.918$
and for this critical reflecting boundary, the zero points of the scalar field
is at the infinity or the scalar field asymptotically approaches zero
according to (5). This possible critical reflecting boundary can be
fixed as the radius of the AdS scalar hairy reflecting star.

\begin{figure}[h]
\includegraphics[width=155pt]{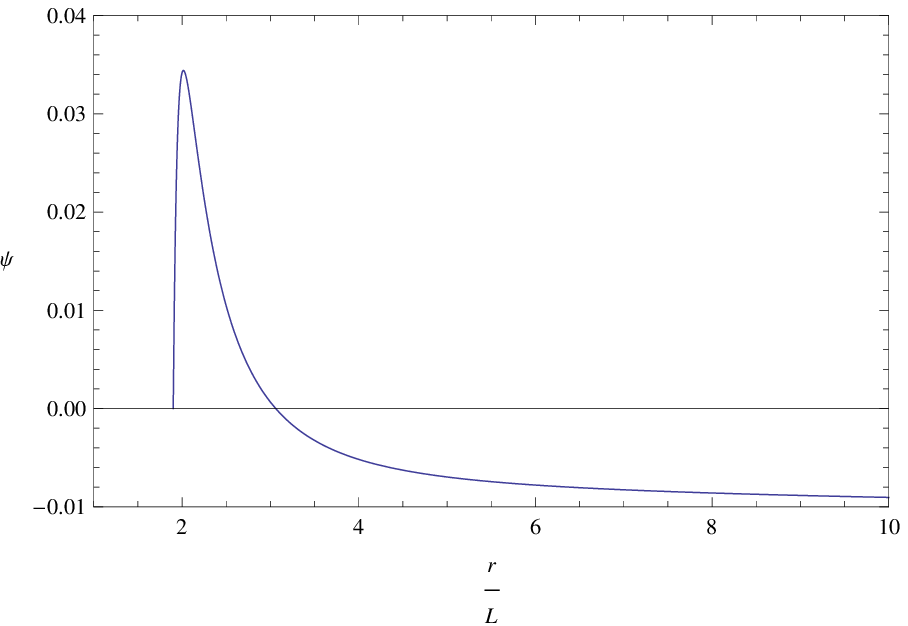}\
\includegraphics[width=155pt]{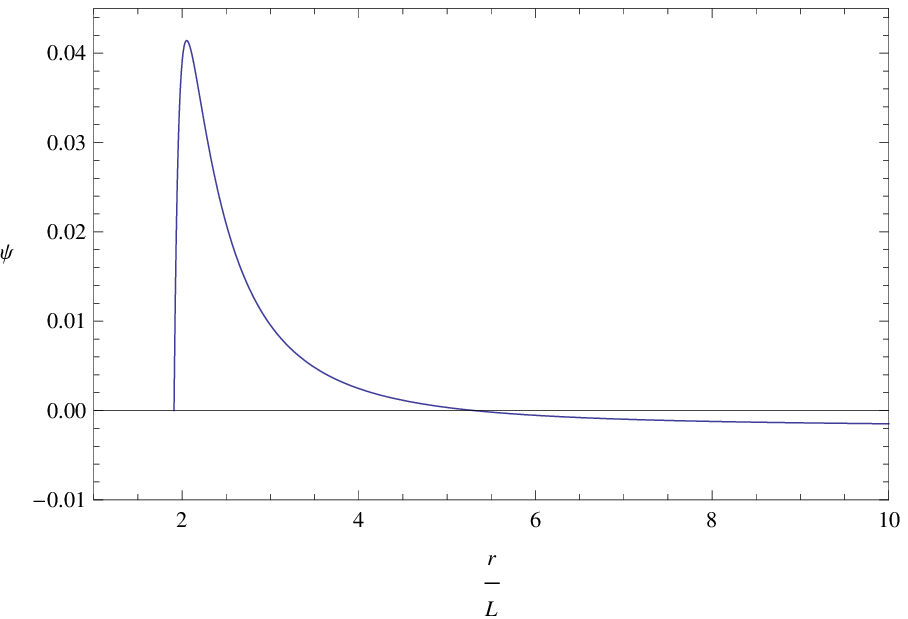}\
\includegraphics[width=155pt]{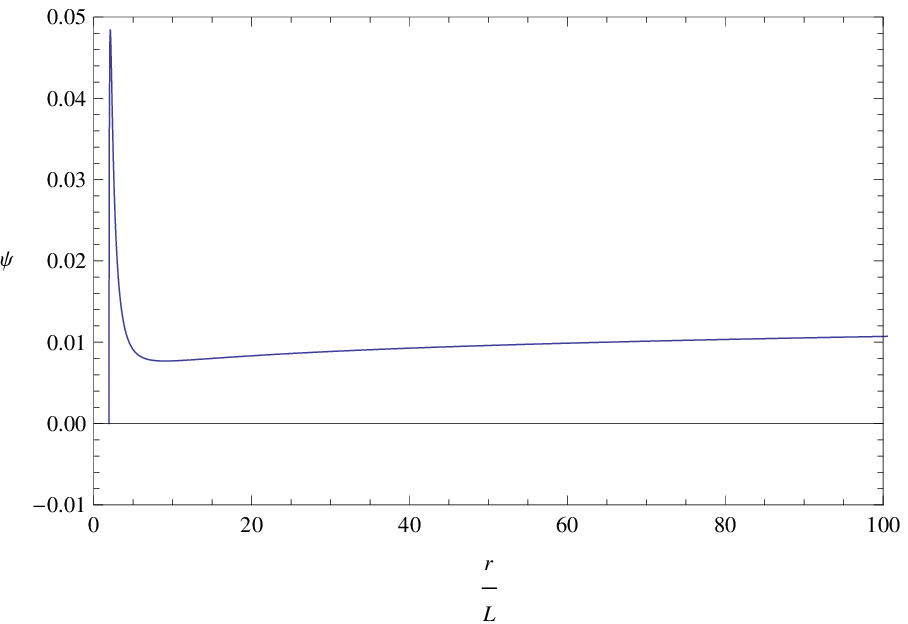}\
\caption{\label{EEntropySoliton} (Color online) We plot the scalar field $\psi$
as a function of the radial coordinate $\frac{r}{L}$ with $qL=2$, $m^2L^2=\frac{1}{2}$, $\frac{Q}{L}=4$, $\frac{M}{L}=7.5$
and various reflecting boundaries. The left panel corresponds to the case
of the reflecting boundary at $\frac{r}{L}=1.900$, the middle panel represents the case of the reflecting boundary at $\frac{r}{L}=1.909$
and the right panel shows the case of the reflecting boundary at $\frac{r}{L}=1.918$.}
\end{figure}

With more detailed calculations in cases of $qL=2$, $m^2L^2=\frac{1}{2}$,
$\frac{Q}{L}=4$ and $\frac{M}{L}=7.5$, we arrive at the critical discrete
reflecting star radius $\frac{r_{s}}{L}\thickapprox1.91077$ with
the scalar field asymptotically approaches zero at the infinity.
As we obtain the discrete reflecting hairy star radius through numerical procedures, we
also analyse the existence of the reflecting radii with other methods.
On the one side, the general solutions of equation (3) behave in the form
$\psi=A\cdot\frac{1}{r^{\lambda_{-}}}+B\cdot\frac{1}{r^{\lambda_{+}}}+\cdots$
with $\lambda_{\pm}=(3\pm \sqrt{11})/2$ as $r\rightarrow \infty$.
We have numerically found $A<0$ for reflecting boundary $\frac{r}{L}$ below 1.91077 and
there is $A>0$ for reflecting boundary $\frac{r}{L}$ above 1.91077 in accordance with results in Fig. 1.
So a critical radius $\frac{r_{s}}{L}\approx 1.91077$ with $A=0$ should exist and the corresponding
scalar field behaves as $\psi\varpropto \frac{1}{r^{(3+\sqrt{11})/2}}$ at the infinity.
On the other side, we have numerically confirmed the natural fact that when
$\frac{r}{L}\rightarrow \frac{r_{s}}{L}\thickapprox1.91077$,
the scalar field with the reflecting boundary imposed at $\frac{r}{L}$
approaches to the scalar field with the reflecting surface putted at
$\frac{r_{s}}{L}\thickapprox1.91077$.

We plot behaviors of the scalar field with the surface at $\frac{r_{s}}{L}=1.91077$
in the case of $qL=2$, $m^2L^2=\frac{1}{2}$, $\frac{Q}{L}=4$ and $\frac{M}{L}=7.5$ in the left panel of Fig. 2.
In fact, $\frac{r_{s}}{L}\approx 1.91077$ is the largest radius of the hairy star
since there is no larger zero points of the scalar field
below the upper bound $\frac{r_{s}}{L}\leqslant max \{2,\sqrt[3]{30},4\}=4$ due to (17).
Integrating the equation from $\frac{r_{s}}{L}=1.91077$ to smaller radial
coordinates in the right panel of Fig. 2, we can obtain various
discrete radii of AdS hairy reflecting stars
around $m r_{s}\thickapprox1.35114,~~1.31465,~~1.31105,~~1.31068$
below the upper bound $m r_{s}\leqslant max \{\sqrt{2},\frac{\sqrt[3]{30}}{\sqrt{2}},2\sqrt{2}\}=2\sqrt{2}
\approx 2.82843$ according to (17).
Here, we find that radii of AdS scalar hairy reflecting stars
are discrete qualitatively the same to analytical results in other reflecting object background \cite{Hod-10,Hod-12,YP2018}.

\begin{figure}[h]
\includegraphics[width=180pt]{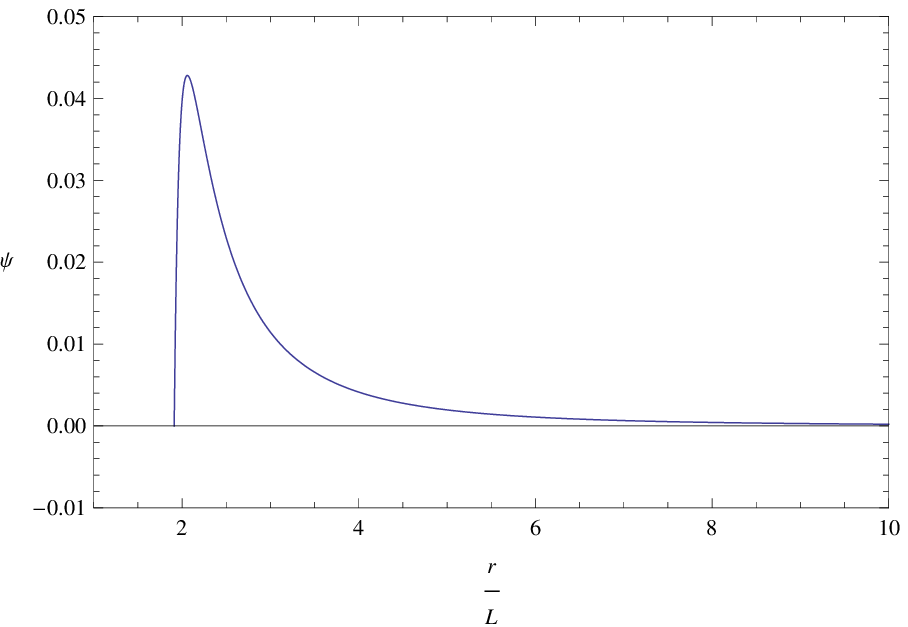}\
\includegraphics[width=180pt]{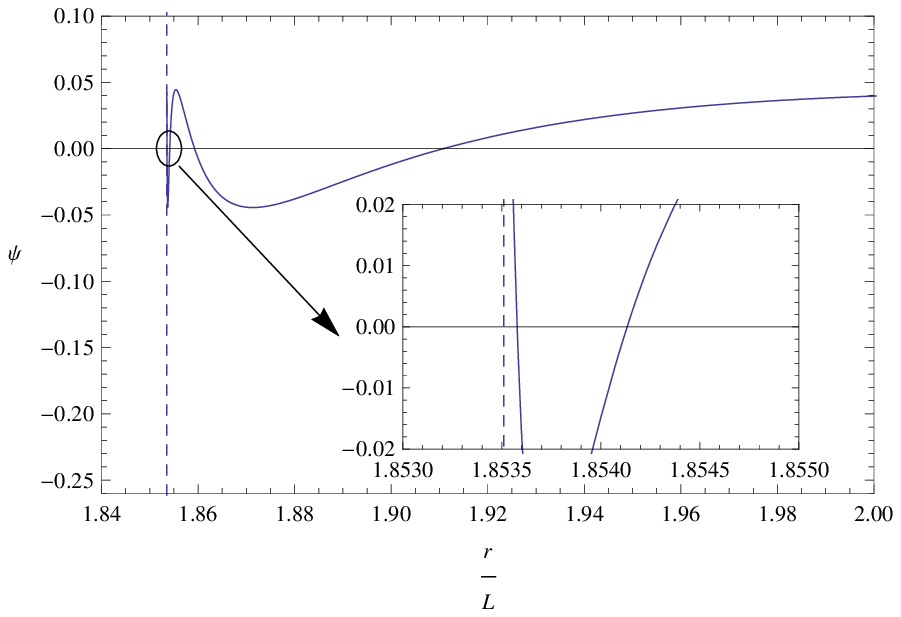}\
\caption{\label{EEntropySoliton} (Color online) We show the scalar field $\psi$
as a function of the radial coordinate $\frac{r}{L}$ with $qL=2$, $m^2L^2=\frac{1}{2}$, $\frac{Q}{L}=4$ and
$\frac{M}{L}=7.5$. The left panel is with the largest radius putted at $\frac{r_{s}}{L}=1.91077$
and the right panel shows various discrete hairy star radii.}
\end{figure}

We numerically obtain four roots of the scalar eigenfunction $\psi$
in the right panel of Fig. 2. Here,we find that
the smallest root $\frac{r_{s}}{L}\thickapprox1.85358$ is very closed to the would be black hole horizon
$\frac{r_{h}}{L}\thickapprox 1.85351$. Since the scalar field changes very quickly around the horizon,
we need more precise method to integrate the equation to smaller radial coordinate.
In fact, it is directly to check the existence of various roots of the scalar field in the AdS reflecting shell with (18) of \cite{YPS},
which implies that there may be also many roots of the scalar field in the AdS reflecting star.

Instead of Dirichlet boundary conditions, we can also impose
Neumann boundary conditions in the form $\psi'(r_{n})=0$
with $r_{n}$ as the surface of the star.
At present, reflecting stars with Neumann surface conditions
have been discussed in \cite{Hod-12,EPV,YPS,Hod-13,Hod-14,YPN}.
We show the scalar field with Neumann reflecting conditions
with $qL=2$, $m^2L^2=\frac{1}{2}$, $\frac{Q}{L}=4$,
$\frac{M}{L}=7.5$ and $\frac{r_{n}}{L}=2.05954$ in Fig. 3.
In fact, the results in Fig. 2 show that there are various discrete extremum points with $\psi'(r)=0$,
which can be fixed as the hairy Neumann reflecting star radius $r_{n}$.

\begin{figure}[h]
\includegraphics[width=220pt]{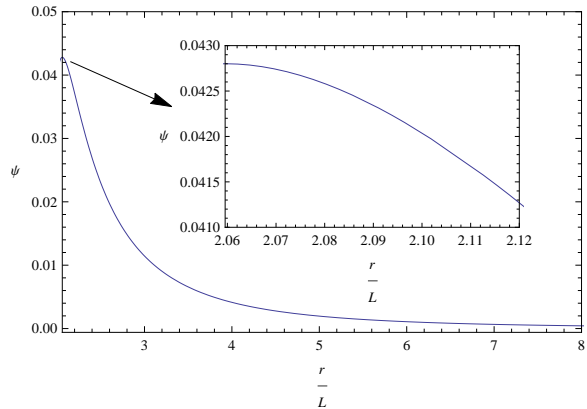}\
\caption{\label{EEntropySoliton} (Color online) We show the scalar field $\psi$
as a function of the radial coordinate $\frac{r}{L}$ with $qL=2$, $m^2L^2=\frac{1}{2}$, $\frac{Q}{L}=4$,
$\frac{M}{L}=7.5$ and the Neumann reflecting star surface putted at $\frac{r_{n}}{L}=2.05954$.}
\end{figure}

We study the largest hairy star radius $m R_{s}$ as a function of
star charge, star mass and cosmological constant with dimensionless quantities according to the symmetry (4) in Fig. 4.
With $qL=2$, $m^2L^2=\frac{1}{2}$ and $\frac{Q}{L}=4$ in the left panel, we see that the
larger star mass $qM$ corresponds to a larger $m R_{s}$
similar to known results in the asymptotically flat reflecting star background \cite{YP2018}.
In contrast, we can see that larger $qQ$ leads to a smaller $m R_{s}$
with $qL=2$, $m^2L^2=\frac{1}{2}$ and $\frac{M}{L}=7.5$ in the middle panel, which is
very different from cases of reflecting stars in the asymptotically flat background \cite{YP2018}.
We also show the functional dependence
of the largest hairy star radius on the value of the cosmological constant $\frac{\Lambda}{q^2}$
with $\frac{m^2}{q^2}=\frac{1}{8}$, $qQ=8$ and $qM=15$ in the right panel.
It can be seen from the picture that the largest radius
decreases as the cosmological constant becomes more negative.
In particular, we find that the largest radius is zero around $\frac{\Lambda}{q^2}\thickapprox -7.322$ and
there is no hairy reflecting star for $\frac{\Lambda}{q^2}\leqslant -7.322$.

\begin{figure}[h]
\includegraphics[width=155pt]{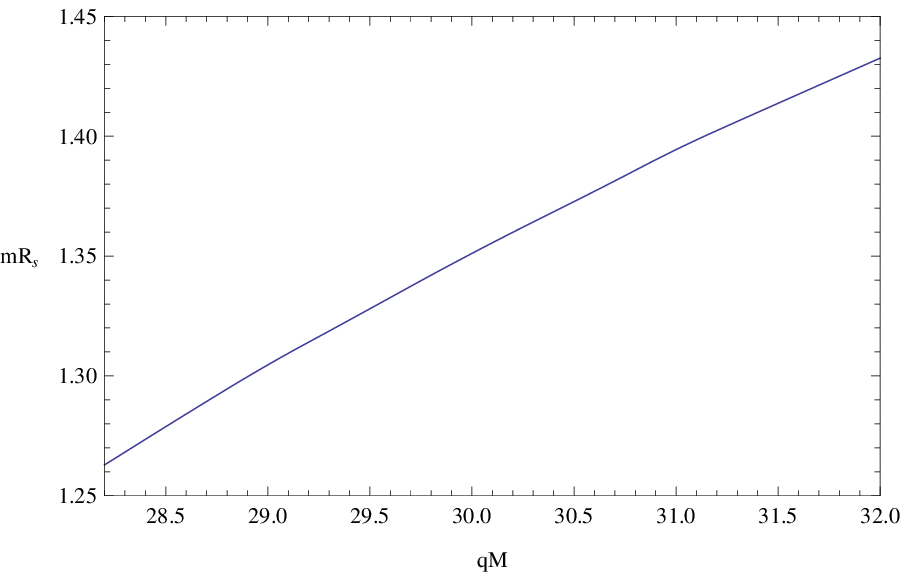}\
\includegraphics[width=155pt]{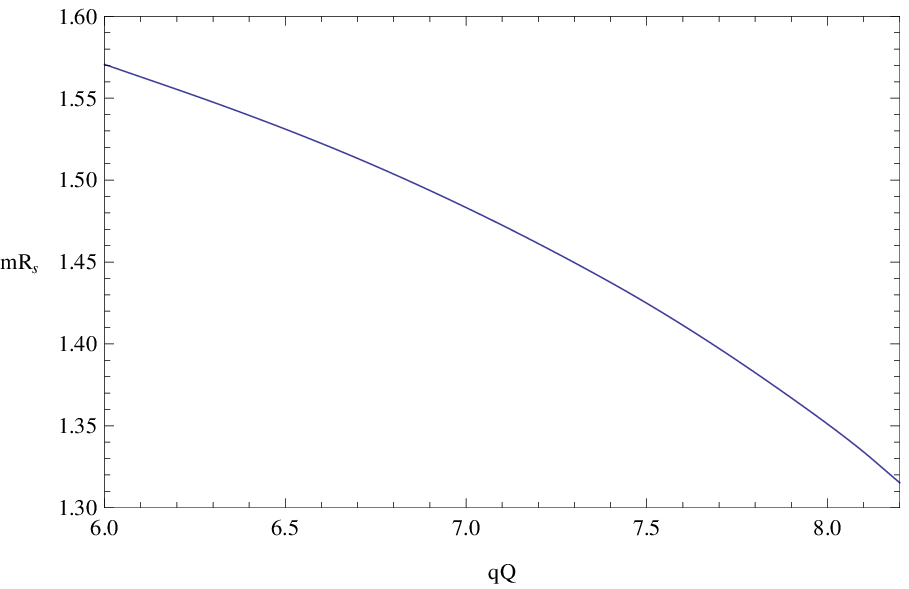}\
\includegraphics[width=155pt]{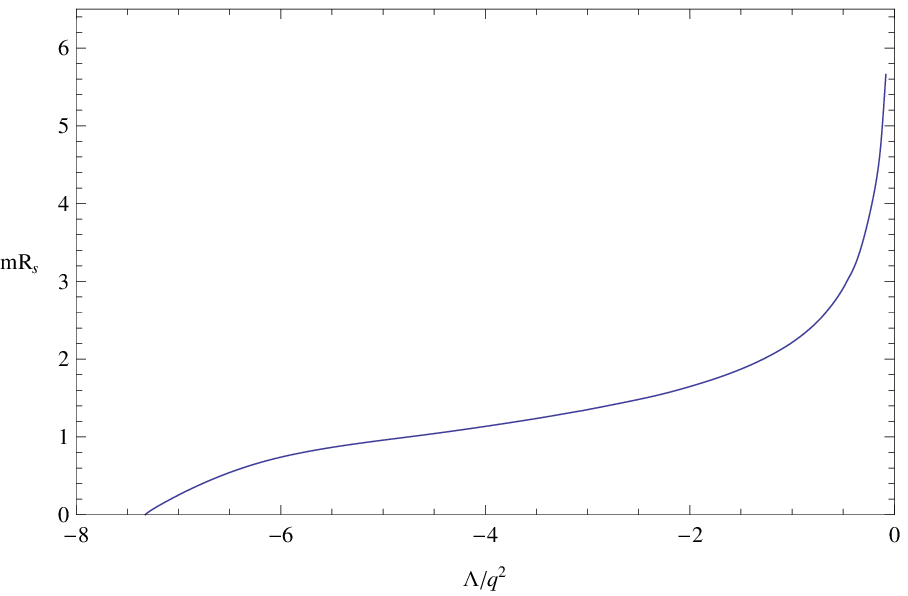}\
\caption{\label{EEntropySoliton} (Color online) We plot the largest hairy star radius with respect to
various star charge, star mass and cosmological constant. We show effects of the star mass $qM$ on $m R_{s}$ with $qL=2$,
$m^2L^2=\frac{1}{2}$ and $\frac{Q}{L}=4$ in the left panel. The middle panel corresponds to behaviors
of $m R_{s}$ with respect to $qQ$ in the case of $qL=2$, $m^2L^2=\frac{1}{2}$ and $\frac{M}{L}=7.5$.
And the right panel is with the largest radius as a function of the cosmological constant $\frac{\Lambda}{q^2}$
in cases of $\frac{m^2}{q^2}=\frac{1}{8}$, $qQ=8$ and $qM=15$.}
\end{figure}

It is interesting to extend the discussion in the middle
panel of Fig. 4 to smaller star charge and also examine
whether there are neutral AdS hairy reflecting stars.
We show the largest star radius $m R_{s}$ and the largest black hole horizon $m R_{h}$
as a function of star charge with $qL=2$, $m^2L^2=\frac{1}{2}$ and $\frac{M}{L}=7.5$
in Fig. 5. Since the reflecting star is assumed to be regular,
the star surface coordinate should be larger than the largest horizon
of a black hole with the same charge and mass.
We find that the largest radius is above the largest horizon in the range $qQ\geqslant4$
and they both increase as we decrease the star charge.
When we choose smaller star charge, the largest star radius becomes nearer to the largest horizon
and for small star charge, they almost coincide with each other.

\begin{figure}[h]
\includegraphics[width=200pt]{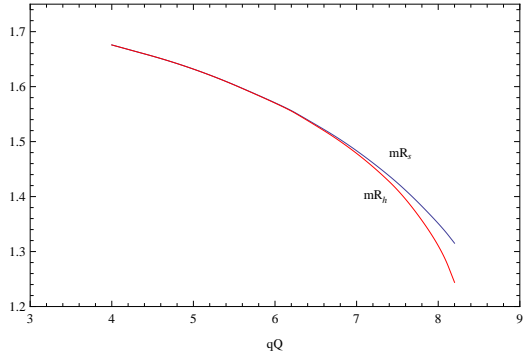}\
\caption{\label{EEntropySoliton} (Color online) We show the largest hairy star radius $m R_{s}$
and largest horizon $m R_{h}$ as a function of
the star charge $qQ$ with $qL=2$, $m^2L^2=\frac{1}{2}$ and $\frac{M}{L}=7.5$.
The top blue line shows behaviors of $m R_{s}$
and the bottom red line is with values of $m R_{h}$.}
\end{figure}

We also show values of the largest star radius $m R_{s}$ and the largest horizon $m R_{h}$ in Table I
with $qL=2$, $m^2L^2=\frac{1}{2}$, $\frac{M}{L}=7.5$ and various $qQ$.
The data again shows that the star surface becomes nearer to the largest horizon
as we decrease the star charge. And when the star charge is $qQ\thickapprox4$, the
star radius is almost equal to the horizon and our further
numerical results suggest that there is no reflecting hairy star with the reflecting surface
outside the horizon for $qQ<4$.
So we conclude that there is no regular asymptotically AdS hairy reflecting star solutions
or no scalar hair theorem holds for small star charge.

\renewcommand\arraystretch{1.7}
\begin{table} [h]
\centering
\caption{The largest radius $m R_{s}$ and largest horizon $m R_{h}$ together with various $qQ$.}
\label{address}
\begin{tabular}{|>{}c|>{}c|>{}c|>{}c|>{}c|>{}c|}
\hline
$~~~~~~~qQ~~~~~~~$ & ~~~~~~~4~~~~~~~ & ~~~~~~~5~~~~~~~& ~~~~~~~6~~~~~~~& ~~~~~~~7~~~~~~~& ~~~~~~~8~~~~~~~\\
\hline
$~~~~~~~m R_{s}~~~~~~~$ & ~~~~~~~1.67585~~~~~~~ & ~~~~~~~1.63194~~~~~~~& ~~~~~~~1.57063~~~~~~~& ~~~~~~~1.48316~~~~~~~& ~~~~~~~1.35112~~~~~~~\\
\hline
$~~~~~~~m R_{h}~~~~~~~$ & ~~~~~~~1.67585~~~~~~~ & ~~~~~~~1.63189~~~~~~~& ~~~~~~~1.57004~~~~~~~& ~~~~~~~1.47850~~~~~~~& ~~~~~~~1.31063~~~~~~~\\
\hline
\end{tabular}
\end{table}

\section{Conclusions}

We studied static scalar field configurations supported by a reflecting star in the AdS spacetime.
With analytical methods, we provided upper bounds for the radius of the scalar hairy star
as $m r_{s}\leqslant  max \{m\sqrt{QL},m\sqrt[3]{4ML^2},\sqrt[4]{2}\sqrt{qQmL}\}$.
Above the bound, the AdS reflecting star cannot support the scalar field or there is no scalar hair theorem.
Below the bound, we numerically obtained the charged scalar hairy reflecting star solutions and
the radius of the hairy star is discrete, which is similar to cases in other reflecting object backgrounds.
For each fixed set of parameters, we searched for the largest radii $mR_{s}$
and also examined effects of parameters on the largest AdS hairy star radii.
We found that larger star mass qM corresponds to a larger $m R_{s}$
similar to cases in the asymptotically flat reflecting star background.
However, we saw that larger star charge qQ leads to a smaller $m R_{s}$,
which is very different from results in asymptotically flat reflecting star spacetime.
We also found that $m R_{s}$ decreases as we choose a more negative cosmological constant.
And our further detailed calculation showed that scalar fields cannot condense
around regular AdS reflecting stars when the star charge is small
or the cosmological constant is negative enough.

\begin{acknowledgments}

This work was supported by the National Natural Science Foundation of China under Grant No. 11305097;
Shandong Provincial Natural Science Foundation of China under Grant No. ZR2018QA008.

\end{acknowledgments}

\end{document}